\documentclass[12pt]{iopart}

\usepackage{iopams}
\usepackage{graphicx,color}
\usepackage{stmaryrd}
\expandafter\let\csname equation*\endcsname\relax
\expandafter\let\csname endequation*\endcsname\relax
\usepackage{mathtools}
\usepackage{multirow}
\usepackage[dvipsnames]{xcolor}
 \usepackage{tikz}
 \usetikzlibrary{chains,shapes,fit,calc,arrows}
\usepackage{pgfplots, pgfplotstable}
\usepgfplotslibrary{groupplots}

\usepackage{color}

\newcommand{\mean}[1]{{\left< #1 \right>}}

\definecolor{webgreen}{rgb}{0,.5,0}
\definecolor{webbrown}{rgb}{.6,0,0}
\definecolor{grigio}{rgb}{.85,.85,.85} 
\definecolor{RoyalBlue}{rgb}{0.0, 0.14, 0.4}
\definecolor{skyblue1}{rgb}{0.45,0.62,0.81}
\definecolor{skyblue2}{rgb}{0.2,0.39,0.64}
\definecolor{skyblue3}{rgb}{0.13,0.29,0.53}
\definecolor{scarlet1}{rgb}{0.93,0.16,0.16}
\definecolor{scarlet2}{rgb}{0.8,0,0}
\definecolor{scarlet3}{rgb}{0.64,0,0}

\usepackage[colorlinks, citecolor = blue, urlcolor = blue]{hyperref}

\begin{document}

\title[Beyond Thermodynamic Uncertainty Relations]{Beyond Thermodynamic Uncertainty Relations: nonlinear response, error-dissipation trade-offs, and speed limits}

\author{Gianmaria Falasco$^{1}$, Massimiliano Esposito$^{1}$, Jean-Charles Delvenne$^{2}$}

\address{$^1$Complex Systems and Statistical Mechanics, Physics and Materials Science Department, University of Luxembourg, L-1511 Luxembourg}

\address{$^2$Institute of Information and Communication Technologies, Electronics and Applied Mathematics Universit\'{e} catholique de Louvain, Louvain-La-Neuve, Belgium}

\ead{gianmaria.falasco@uni.lu}
\ead{massimiliano.esposito@uni.lu}
\ead{jean-charles.delvenne@uclouvain.be}


\begin{abstract}

From a recent geometric generalization of Thermodynamic Uncertainty Relations (TURs) we derive novel upper bounds on the nonlinear response of an observable of an arbitrary system undergoing a change of probabilistic state. Various relaxations of these bounds allow to recover well known bounds such as (strengthenings of) Cramer-Rao's and Pinsker's inequalities. 
 In particular we obtain a master inequality, named Symmetric Response Intensity Relation, which recovers several TURs as particular cases. 
We employ this set of bounds for three physical applications. First, we derive a trade-off between thermodynamic cost (dissipated free energy) and reliability of systems switching instantly between two states, such as one-bit memories. We derive in particular a lower bound of $2.8 k_BT$ per Shannon bit to write a bit in such a memory, a bound distinct from Landauer's one. 
Second, we obtain a new family of classic speed limits which provide lower bounds for non-autonomous Markov processes on the time needed to transition between two probabilistic states in terms of a thermodynamic quantity (e.g. non-equilibrium free energy) and a kinetic quantity (e.g. dynamical activity). 
Third, we provide an upper bound on the nonlinear response of a system based solely on the `complexity' of the system (which we relate to a high entropy and non-uniformity of the probabilities). We find that `complex' models (e.g. with many states) are necessarily fragile to some perturbations, while simple systems are robust, in that they display a low response to arbitrary perturbations.

\end{abstract}

\maketitle

\section{Introduction}
 
Thermodynamic Uncertainty Relations (TURs) are as old as their better known counterparts of quantum mechanics.
For more than a half century the term has denoted the tradeoff between the uncertainty in pairs of conjugated thermodynamic variables in systems at equilibrium (see \cite{uffink1999thermo} and references therein). Despite a long debate about their exact practical meaning, their derivation has been repeated several times, resorting either to the theory of equilibrium fluctuations or to purely statistical methods \cite{puglisi2017temp}.

However, in the last few years the term has started to designate a radically different set of inequalities valid for nonequilibrium systems (see \cite{bar15a, GingHorEng16, pol16, hor17, pro17, gin17fp, dechant18current, potts2019thermodynamic, hasegawa19, van20, liu20, falasco2020unifying} for an incomplete list).  Roughly speaking, TURs state that the mean-to-variance ratio of current-like observables is bounded by a function of the entropy production \cite{hor19}.
These TURs have been obtained by analogous methods as those used for the ancient thermodynamics uncertainty relations, in their modern forms now well known to physicists---large deviations theory \cite{GingHorEng16,hor17} and information theory \cite{dechant2020fluctuation}---and applied to rather general dynamics---Markov processes equipped by the thermodynamic notion of local detailed balance \cite{van15}.
Recently, we have put forward an alternative derivation which deduces TURs from the very mathematical properties of the observable space \cite{falasco2020unifying}---somewhat similar to viewing the Heisenberg principle as a direct result of the uncertainty between Fourier transformed pairs of variables.

The fact that nonequilibrium conditions bring time and dynamics to the forefront has led the community to consider other bounds on dissipative processes. On the one hand, classical speed limits (SLs) have been derived which prescribe the minimal time to implement a system's  transformation in terms of the involved dissipation  \cite{shi18}. It is by now understood that TURs and SLs are just two faces of the same fundamental fact underlying all nonequilibrium conditions \cite{vo20}, that is the breaking of time-reversal invariance and hence the appearance of a non-zero average entropy production. On the other hand, results have started to appear on the response of systems to environmental perturbations \cite{Owen20}.
The general picture emerging from these various results is the identification of the fundamental physical tradeoffs between efficiency, performance and robustness of nonequilibrium processes, as used in natural as well as man-made applications.

In this paper we contribute to the quest for a unifying theory.
First we define the Asymmetric Response Intensity of an observable $f$ subject to a change of probability distribution from $q$ to $p$. We recover the linear regime bound of \cite{dechant2020fluctuation} but also a nonlinear version, which turns out to coincide with a well-known generalization of Cramer-Rao's relation. Then we introduce the Symmetric Response Intensity, a related quantity that nevertheless obeys different inequalities. In particular we derive the  Symmetric Response Intensity Relation, that contains the various TURs for anti-symmetric observables. Remarkably, it allows to derive a relation between reliability and thermodynamic cost for memories or gates that relax abruptly from a zero to a one conversely (thus dissipating entropy on the way). This model is not universal (as one can escape this relation by a slow protocol for instance) but is relevant for many current technologies using switched systems, even beyond computing devices. As a further application, we also recover a result on distinguishing the arrow of time, completing \cite{rol15}.

The maximum response intensities between two probability distributions can serve as a measure of distance or divergence between these distributions. We leverage this to obtain novel SLs \cite{shi18, vo20} for finite state Markov chains, which offer lower bounds on the time needed to reach a probability distribution from another, in terms of the distance of these probability distribution, the distance to stationarity and the maximum escape (or entrance) rates.

 Finally we observe that the largest response one can obtain away from a probability distribution is still relatively small if this probability is `simple', i.e. of low entropy and close to uniform distribution. Thus `simple' systems are `robust'. In particular, one should be cautious in using simplistic toy models to model complex situations (a practice often found in biophysics), as these toy models may be much more robust to perturbations than the original systems.
 
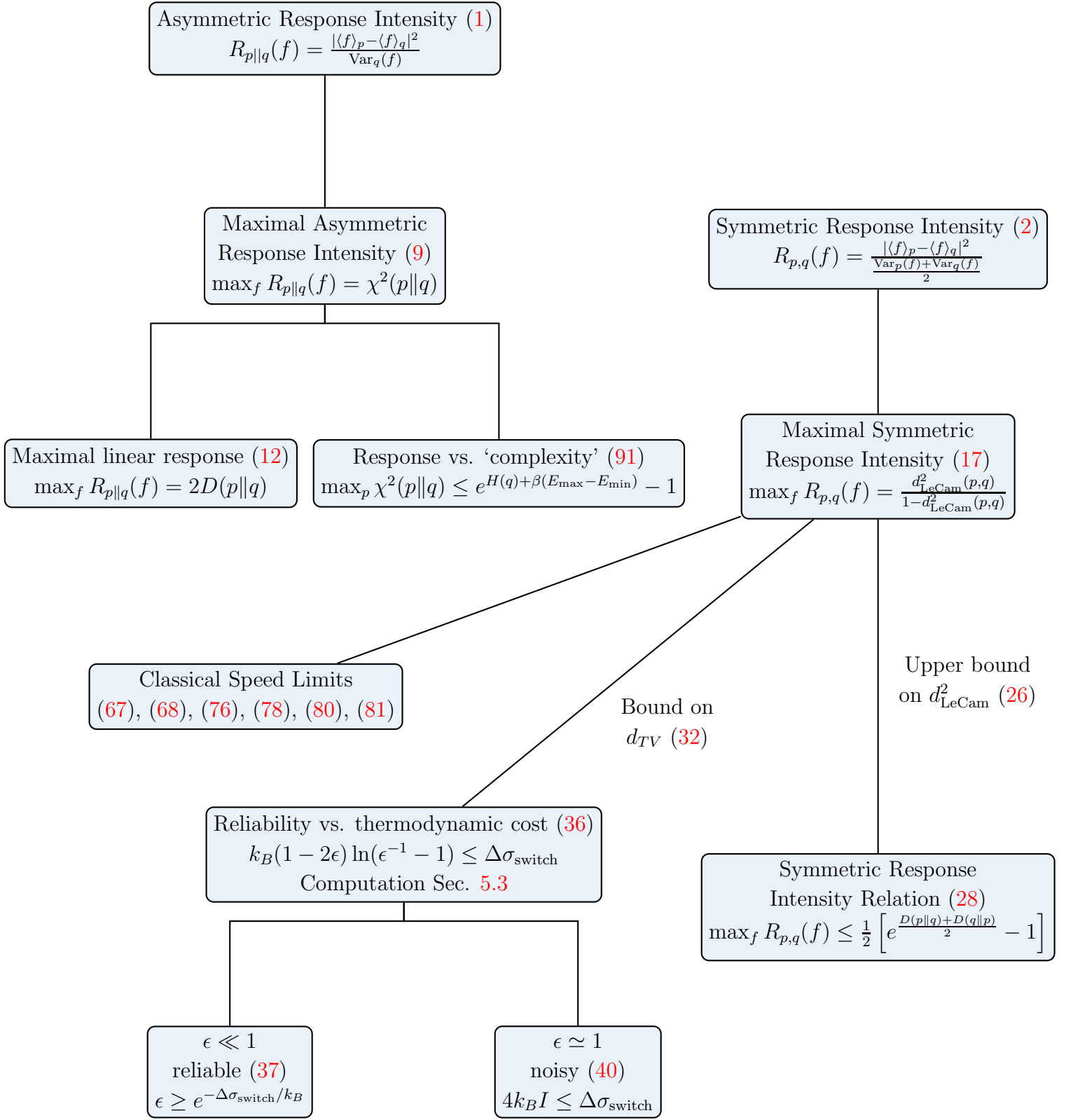
\begin{figure}[htbp]
\hspace{-9.3cm}
\tikzstyle{every picture}+=[remember picture]
\begin{tikzpicture}[sibling distance=16em, level distance=10em,
thick,edge from parent/.style = {draw, edge from parent path={(\tikzparentnode.south)
-- +(0,-10pt)   -| (\tikzchildnode)}}, treenode/.style = {shape=rectangle, rounded corners, draw, align=center, top color=skyblue1!15, bottom color=skyblue1!15}]

\node[treenode] {Asymmetric Response Intensity   \eqref{eq:def-asym-resp-int} \\
$R_{p||q}(f)=\frac{|\langle f \rangle_p - \langle f \rangle_q|^2}{\text{Var}_q(f)}$}
child { node[treenode] (l1) {Maximal Asymmetric\\ Response Intensity  \eqref{eq:HCRbound}\\
$\max_f R_{p\|q}(f) = \chi^2(p\|q)$  }
child{ node[treenode] (l3) {Maximal linear response  \eqref{eq:linear_resp}  \\ $\max_f R_{p\|q}(f) = 2D(p\|q) $ }}
child{ node[treenode] (l4) { Response vs. `complexity'  \eqref{eq:qmin} \\ $ \max_p \chi^2(p\|q)  \leq   e^{H(q)+\beta(E_{\max}-E_{\min}) }-1 $ 
 } }};

\node[treenode] at (10.5,-4) {Symmetric Response Intensity  \eqref{eq:def-sym-resp-int} \\
$R_{p,q}(f)=\frac{|\langle f \rangle_p - \langle f \rangle_q|^2}{\frac{\text{Var}_p(f)+\text{Var}_q(f)}{2}} $}
child { node[treenode] (l5) {Maximal Symmetric\\ Response Intensity  \eqref{eq:maxRsym} \\
$\max_f R_{p,q}(f)=\frac{d^2_\text{LeCam}(p,q)}{1-d^2_\text{LeCam}(p,q)}$ }};

%

\node at (-13,-6.5) {$p \approx q$};
\node at (-8,-6.5) {\begin{tabular}{c}  Finite\\ space \end{tabular}};

\node at (12.2,-12.2) {\begin{tabular}{c} Upper bound \\ on $d^2_\text{LeCam}$ \eqref{eq:mainboundR0dKL} \end{tabular}};

\node[treenode] (r1) at (-1.5,-12.5) {Classical Speed Limits \\  \eqref{eq:LeCamspeedlimit},  \eqref{eq:LeCamkinetic}, \eqref{eq:CSLTVTV}, \eqref{eq:CSLTVTV_pinsker},  \eqref{eq:CSLTVTV_BH}, \eqref{eq:CSLTVTV_kin}};
\draw (l5) edge node [sloped] {} (r1);

\node[treenode] (r2) at (10.5,-16.5) { Symmetric Response\\ Intensity Relation \eqref{eq:Rpqsym-KL}\\
$ \max_f R_{p,q}(f) \leq  \frac{1}{2} \left[e^{\frac{D(p\|q) + D(q\|p)}{2}}-1 \right] $ };
\draw (l5) edge node [sloped] {} (r2);

\node[treenode] (r3) at (1.5,-15.5){ Reliability vs. thermodynamic cost \eqref{eq:eps-sigma}\\
$k_B (1-2 \epsilon) \ln(\epsilon^{-1}-1) \leq \Delta \sigma_\text{switch}$ \\
Computation Sec. \ref{sec:computation} }
child { node[treenode] (l8) {$\epsilon \ll 1$\\ reliable \eqref{eq:error_reliable}\\ $\epsilon \geq e^{-\Delta \sigma_\text{switch}/k_B}$ }}
child { node[treenode] (l9) {$\epsilon \simeq 1$\\ noisy  \eqref{eq:I-sigma} \\ $4 k_B I \leq \Delta \sigma_\text{switch}$}};

\draw (l5) edge node [sloped] {} (r3);

\node at (6.5,-13) {\begin{tabular}{c} Bound on \\ $d_{TV}$ \eqref{eq:d-vs-KL-ln} \end{tabular}};

		\end{tikzpicture}
	\caption{ Graphic outline of the main results of the paper and their logical connection.}
	\label{fig1}
\end{figure}

 To guide the reader though the text we collected the main results in the graph Figure~\ref{fig1} and the most important quantities in Table~\ref{tab1}.

\begin{table}[]
\center
\begin{tabular}{|c|c|}
\hline
\begin{tabular}[c]{@{}c@{}}Kullback-Leibler divergence \end{tabular}               & \begin{tabular}[c]{@{}c@{}} $D(p \| q) = \sum_{\omega \in \Omega} p(\omega) \log \frac{p(\omega)}{q(\omega)} $ \end{tabular} \\ \hline
\begin{tabular}[c]{@{}c@{}} $\chi^2$ divergence \end{tabular}   & \begin{tabular}[c]{@{}c@{}} $\chi^2(p\|q)= \sum_{\omega \in  \Omega} \frac{p^2(\omega)}{q(\omega)}-1$  \end{tabular}     \\ \hline
\begin{tabular}[c]{@{}c@{}} Le Cam's distance \end{tabular}   & {$d_\text{LeCam}(p,q)=\sqrt{\frac{1}{2}\sum_{\omega \in  \Omega} \frac{(p(\omega)-q(\omega))^2}{p(\omega)+q(\omega)}}  $  }                                                                                                                                                                                                                                                                                                              \\ \hline
\begin{tabular}[c]{@{}c@{}} Shannon entropy \end{tabular} & \begin{tabular}[c]{@{}c@{}} $H(q)= - \sum_{\omega \in  \Omega} p(\omega) \log p(\omega) $ \end{tabular}                             \\ \hline
\begin{tabular}[c]{@{}c@{}} Total variation distance \end{tabular}               & {$d_{TV}(p,q)=\frac{1}{2} \|p-q\|_1 = \frac{1}{2} \sum_{\omega \in \Omega} |p(\omega)-q(\omega)|$   }                                                                                                                                                                                                                                                                                    \\ \hline
\end{tabular}
\caption{Summary of the main quantities used in the paper.}
\label{tab1}
\end{table}

\section{How to measure nonlinear response}

Consider two possible probability distributions $p$ and $q$ on an arbitrary space $\Omega$. A given physical observable, i.e. a real random variable $f: \Omega \to \mathbb{R}$ may exhibit two different expected values $\langle f \rangle_p$ and $\langle f \rangle_q$. In order to check how significant the difference between the two is, we may compare it to the average fluctuations of the observable under $p$ or $q$. Taking $q$ as a `reference', or `unperturbed' probability distribution, and $p$ as the `perturbed' distribution, we see  the following adimensional ratio as the \emph{response intensity} of the observable $f$ to the perturbation $p$ of $q$:
\begin{equation}\label{eq:def-asym-resp-int}
R_{p||q}(f)=\frac{|\langle f \rangle_p - \langle f \rangle_q|^2}{\text{Var}_q(f)}.
\end{equation}
We call it the \emph{asymmetric} response intensity as $q$ and $p$ have distinct roles.
Sometimes $p$ and $q$ are two probability distributions of interest to describe a situation, without hierarchy of role. For example, think at two metastable states realizable within the same physical system.
In this situation we may simply consider the average fluctuations over $p$ and $q$, and define the \emph{symmetric response intensity}:

\begin{align}\label{eq:def-sym-resp-int}
R_{p,q}(f)=\frac{|\langle f \rangle_p - \langle f \rangle_q|^2}{\frac{\text{Var}_p(f)+\text{Var}_q(f)}{2}}.
\end{align}

One may  intuitively see these quantities as how easy it is to tell $p$ and $q$ apart by just observing the value of $f$. If the response intensity $R_{p||q}(f)$ is less than one, then it means that the typical values taken by $f$ under $p$ or under $q$ differ by less than the typical fluctuation of $f$, hence it is hard to conclude whether  the system is in the  distribution $p$ or  $q$. Conversely, a large intensity means that the difference in response can hardly be confused with a random fluctuation.

The response intensity can be likened to a signal-to-noise ratio, if we see $\langle f \rangle_p$ or  $\langle f \rangle_q$ as the signal being transmitted to the environment by a system switching between distributions $p$ and $q$. This analogy can be made formal in the following scenario: if we choose to apply $p$ or $q$ randomly (with equal probabilities), then the signal, taking randomly the value $\langle f \rangle_p$ or $\langle f \rangle_q$, has a variance $|\langle f \rangle_p -  \langle f \rangle_q|^2/4$, thus a signal-to-noise ratio (which is defined  in information theory as the variance of the signal over the variance of the noise) $R_{p,q}(f)/4$.



\section{Link between symmetric and asymmetric response intensities}

We make some elementary observations on the response intensities introduced above.
The symmetric response intensity is the harmonic average of asymmetric response intensities:
\begin{align}
\frac{2}{R_{p,q}(f)}=\frac{1}{R_{p||q}(f)}+\frac{1}{R_{q||p}(f)}
\end{align}
A key observation is that adding a constant to the observable $f$ does not change $R_{p||q}(f)$ or $R_{p,q}(f)$, since both the response $\langle f \rangle_{p} - \langle f \rangle_{q}$ and the variances are unaffected. Thus we can assume without loss of generality that  $\langle f \rangle_{p} = - \langle f \rangle_{q}$. We call such an observable \emph{centered}. 
Let us introduce an intermediate quantity where the variance is replaced with the second moment:
\begin{equation} \label{eq:R0pq}
R^{0}_{p,q}(f)= \frac{|\langle f \rangle_p - \langle f \rangle_q|^2}{\frac{\langle |f|^2 \rangle_p + \langle |f|^2 \rangle_q}{2}}
\end{equation}
For a centered observable $f$, we find by direct calculation that 
\begin{equation}\label{eq:RvsR0}
R_{p,q}(f) / 4 = \frac{R^{0}_{p,q}(f) / 4}{1 - R^{0}_{p,q}(f) / 4}= \frac{1}{\frac{1}{R^{0}_{p,q}(f) / 4}-1}
\end{equation}
and that
\begin{equation} \label{eq:R0pq4}
R^{0}_{p,q}(f)/4 = R_{p||\frac{p+q}{2}}(f)= R_{q||\frac{p+q}{2}}(f).
\end{equation}
We would like now to find a tight upper bound on the response intensities $R_{p,q}(f)$ and $R_{p||q}(f)$, which would hold for all $f$, only depending on $p$ and $q$.  

\section{Tight bounds on nonlinear response intensity and uncertainty principles}

We exploit here the similar form of the response intensity and the square-mean-to-variance ratio appearing in TURs to derive bounds on the former in terms of physically interpretable quantities (such as the Kullback-Leibler divergences) between $p$ and $q$.
Specifically we use a recent geometric formalism introduced under the name of Hilbert Uncertainty Relation \cite{falasco2020unifying} which allows to prove general inequalities including, but not limited to, TURs, as we now recall. 

\subsection{The Hilbert Uncertainty Relation: general statement}

We consider a Hilbert space $\mathcal{H}$ of observables, endowed with an arbitrary scalar product  $\langle . | . \rangle$. One should not confuse $\langle . | . \rangle$, a standard notation for a scalar product, with $\langle . \rangle$, a standard notation introduced above for the mean value (with respect to some probability distribution).
 Given a real continuous linear form $f \mapsto L(f) \in \mathbb{R}$, we want to find the maximum value of the ratio
\begin{equation}\label{eq:ratio-HUR}
\frac{|L(f)|^2}{\langle f | f \rangle}
\end{equation}
among all $f$ belonging to a given closed linear subspace $\mathcal{F} \subseteq \mathcal{H}$ of observables of interest.  We find that it reaches the maximum value for $f=m \in \mathcal{F}$ such that  $L(g)=\langle m | g \rangle$ for all $g \in \mathcal{F}$, with $m$ existing by virtue of Riesz's representation theorem.
Then the maximum ratio takes the value
\begin{align}
\max_f \frac{|L(f)|^2}{\langle f | f \rangle} = \langle m | m \rangle=|L(m)|.
\end{align}
Hereafter the Hilbert $\mathcal{H}$ space can be finite or infinite-dimensional. As Applications II and (part of) III below regard discrete systems, we spell out the general theory by using discrete summation notations, even though all results can be extended in straightforward way to continuous probability distributions, where the sums must in general be replaced with integrals.

\subsection{The Hilbert Uncertainty Relation: application to asymmetric response intensity}

We can immediately apply the Hilbert Uncertainty Relation to derive the observables with maximal response intensity.

Let us start with the asymmetric case: given probability distributions $p$ and $q$, we want to find $f$ that maximises $R_{p||q}(f)$. To apply the Hilbert Uncertainty Relation, we consider the linear form  $L(f) = \langle f \rangle_p - \langle f \rangle_q$ and the scalar product $\langle f | g \rangle = \langle f  g \rangle_q$ (the expected product $fg$ according to distribution $q$). Since, as observed above, we can always shift an observable by a constant and preserve the asymmetric response intensity $R_{p||q}(f)$, we limit ourselves to the observables with zero $q$-mean: $\mathcal{F} = \{f: \Omega \to \mathbb{R} | \langle f   \rangle_q =0  \}$. In this space,  variance and second moment coincide, thus the asymmetric response intensity \eqref{eq:def-asym-resp-int} is indeed the ratio \eqref{eq:ratio-HUR}. We easily find that the optimal observable is $m=\frac{p}{q}-1$. Indeed we can check that $\langle m   \rangle_q =\sum_{\omega \in  \Omega} q(\omega) (\frac{p(\omega)}{q(\omega)}-1)=0$ and $\langle m | g \rangle = \sum_{\omega \in  \Omega} q (\frac{p}{q}-1)g = L(g)$.

Thus we conclude
\begin{equation}\label{eq:HCRbound}
	\max_f R_{p\|q}(f) = \sum_{\omega \in  \Omega} \frac{p^2}{q}-1= \chi^2(p\|q), 
\end{equation}
where $\chi^2(p\|q)$ is the so-called $\chi^2$ divergence of $p$ with respect to $q$.
Equation \eqref{eq:HCRbound} formally coincides with the \emph{Hammersley-Chapman-Robbins bound}  \cite{lehmann2006theory} but has the merit of giving physical meaning to a result of estimation theory, by identifying in the lefthand side of \eqref{eq:HCRbound} a sound proxy for the relative response intensity of a physical system.
The latter bound is a known generalisation of Cramer-Rao's bound \cite{lehmann2006theory}, which is obtained from \emph{Hammersley-Chapman-Robbins bound} by considering $p$ close to $q$, such that the $\chi^2$ divergence reduces to Fisher's information metric. 

The $\chi^2$ divergence has the following relation with the Kullback-Leibler divergence, resulting from convexity of exponential:

\begin{align}\label{eq:chivsKL}
\chi^2(p\|q) \geq e^{D(p\|q)}-1 \geq D(p\|q).
\end{align}
This inequality can be interpreted physically in the following way. If we consider $q$ as a stationary distribution of some Markov process and $p$ as a non-stationary distribution relaxing to $q$, then  $k_B D(p\|q)$ is the entropy produced in the course of the relaxation, not counting the `housekeeping' entropy  possibly produced to maintain the stationary state $q$, if it is not an equilibrium distribution \cite{hat01} (see Section \ref{sec:bit} for a more thorough discussion). This entropy production corresponds exactly to the \emph{non-adiabatic entropy production} \cite{esposito2010three} for the case of autonomously relaxing systems.
 In the case where the stationary state $q$ is an equilibrium at temperature $T$, then $D(p\|q) k_BT$ is the non-equilibrium free energy being dissipated in the course of relaxation \cite{esposito2011second, sivak2012near,par15}.

Thus if $p$ is far from the stationary distribution $q$ (in the sense of a large $D(p\|q)$), then it results in the existence of a large nonlinear response intensity $R_{p\|q}(f)$ for \emph{some} observable $f$. 
In the limit where $p$ and $q$ are close (the so-called \emph{linear response regime}), a Taylor expansion yields
\begin{align}
\chi^2(p\|q)\approx \sum_{\omega \in  \Omega} \frac{(p-q)^2}{q} \approx 2 D(p\|q).
\end{align}
In summmary, we have in the limit of linear regime:
\begin{align}\label{eq:linear_resp}
\max_f R_{p\|q}(f) = 2D(p\|q)
\end{align}
which provides another proof of Dechant-Sasa's linear-regime relation \cite{dechant2020fluctuation}. 

\subsection{The Hilbert Uncertainty Relation: application to symmetric response intensity} \label{sec:sym_resp}

As we seek now to bound the symmetric response intensity, we choose the scalar product $\langle f | g \rangle = \frac{\langle f g \rangle_p +  \langle f g \rangle_q}{2} =\langle f g \rangle_{\frac{p+q}{2}}$. The space of observables $\mathcal{F}$ of interest is now the set of all observables. 
The linear form is still $L(f)=\langle f \rangle_p -  \langle f \rangle_q$. In this case we see that the ratio \eqref{eq:ratio-HUR} is exactly $R^0_{p,q}(f)$, defined in \eqref{eq:R0pq}. 
The maximum ratio is reached for the observable $m=2\frac{p-q}{p+q}$ in application of the Hilbert Uncertainty Relation,  and $\langle m | g \rangle = L(g)$ for any  observable $g$. Thus we arrive at
\begin{equation}\label{eq:R0LeCam}
	\max_{f} R^0_{p,q}(f) = 2\sum_{\omega \in  \Omega} \frac{(p-q)^2}{p+q}=4 d^2_\text{LeCam}(p,q).
\end{equation}

The latter expression involves Le Cam's distance \cite{le2012asymptotic,vincze1981concept}  between two probability distributions $p$ and $q$, defined as 
\begin{equation}\label{eq:defsLeCam}
d_\text{LeCam}(p,q)=\sqrt{\frac{1}{2}\sum_{\omega \in  \Omega} \frac{(p-q)^2}{p+q}}\leq 1.
\end{equation}
This quantity is indeed a proper distance for the space of probability distributions over $\Omega$, verifying the axioms of a metric (including, non-trivially, the triangle inequality) \cite{le2012asymptotic}. 

Note that $m$ happens to be centered, namely $\langle m\rangle_p = - \langle m\rangle_q$, thus the maximum in \eqref{eq:defsLeCam} is also the maximum of all centered observables $f$:

\begin{equation}\label{eq:R0LeCam2}
\max_{f \text{centered}} R^0_{p,q}(f) = 2\sum_{\omega \in  \Omega} \frac{(p-q)^2}{p+q}=4 d^2_\text{LeCam}(p,q).
\end{equation}

In view of \eqref{eq:R0pq4}, the same observable $m=2\frac{p-q}{p+q}$ is also maximal for $R_{p\|\frac{p+q}{2}}(f)$ and for $R_{p\|\frac{p+q}{2}}(f)$---over \emph{all} observables $f$, centered or not, since those quantities are invariant under shift by a constant. Thus from the Hammersley-Chapman-Robbins  bound \eqref{eq:HCRbound}, we find that 
\begin{align}
d^2_\text{LeCam}(p,q)=\chi^2 \left (p \bigg{\|}\frac{p+q}{2} \right)=\chi^2 \left (q\bigg{\|}\frac{p+q}{2} \right),
\end{align}
 as can also be verified by direct calculation. 
 Finally, using \eqref{eq:RvsR0} and \eqref{eq:R0pq4},  the symmetric response intensity is found to be bounded (for \emph{all} observables $f$, centered or not) as follows:
\begin{align} \label{eq:maxRsym}
 \max_f R_{p,q}(f)  &= \frac{\frac{1}{2}\sum_{\omega \in  \Omega} \frac{(p-q)^2}{p+q}}{1-\frac{1}{2}\sum_{\omega \in  \Omega} \frac{(p-q)^2}{p+q}}\\&=\frac{\chi^2(p\|\frac{p+q}{2})}{1-\chi^2(p\|\frac{p+q}{2})}\\&=\frac{d^2_\text{LeCam}(p,q)}{1-d^2_\text{LeCam}(p,q)}
\end{align} 
Understanding the symmmetric response intensity thus reduces to understanding Le Cam's distance, e.g.  through bounds with respect to other quantities such Kullback-Leibler divergences, as we now proceed to do.

\subsection{Upper bound on Le Cam's distance and the Symmetric Response Intensity Relation}

Upper and lower bounds can be found on Le Cam's distance in terms of other distances between probability distributions (see  \ref{app:lower} for lower bounds). A well-known distance between probability distributions is the \emph{total variation distance} 
\begin{align}
d_{TV}(p,q)=\frac{1}{2} \|p-q\|_1 = \frac{1}{2} \sum_{\omega \in \Omega} |p-q|=\sum_{\omega :  p>q} (p-q) \leq 1.
\end{align}
We now derive a novel upper bound on  Le Cam's distance based on the total variation distance and the Kullback-Leibler divergence.
 Given the distributions $p,q$ on $\Omega$, we define $\Omega_p = \{ \omega \in \Omega | p(\omega) > q(\omega) \}$ and $\Omega_q = \{ \omega \in \Omega | q(\omega) \geq  p(\omega) \}$. In other words, we split $\Omega$ into the part $\Omega_p$ (resp. $\Omega_q$) where $p$ dominates $q$ (resp. where $q$ dominates $p$). On $\Omega_p$ we define the distribution $p'=\frac{p-q}{d_{TV}(p,q)}$, while on $\Omega_q$ we define the distribution $q'=\frac{q-p}{d_{TV}(p,q)}$.
We now observe that $\frac{p-q}{p+q}=\tanh \frac{1}{2}\ln \frac{p}{q}$ and exploit the fact that $\tanh$ is concave for positive arguments.
In this way we find
\begin{align}
\sum_{\omega \in \Omega}& \frac{(p-q)^2}{p+q} = \sum_{\omega \in \Omega} (p-q) \tanh \frac{1}{2} \ln \frac{p}{q}\\
&= d_{TV}(p,q) (\sum_{\omega \in \Omega_p} p'  \tanh \frac{1}{2}\ln \frac{p}{q} \quad + \quad  \sum_{\omega \in \Omega_q} q'  \tanh \frac{1}{2}\ln \frac{q}{p})\\
&\leq  d_{TV}(p,q) ( \tanh \frac{1}{2}  \sum_{\omega \in \Omega_p} p' \ln \frac{p}{q} \quad + \quad  \tanh \frac{1}{2}  \sum_{\omega \in \Omega_q} q' \ln \frac{q}{p}   )\\
&\leq 2  d_{TV}(p,q)  \tanh \frac{1}{4 d_{TV}(p,q)}  \sum_{\omega \in \Omega} (p-q) \ln \frac{p}{q} \\
&= 2d_{TV}(p,q)  \tanh \frac{1}{2 d_{TV}(p,q)}  \frac{D(p\|q)  +  D(q\|p)}{2}
\end{align}
To summarize, our first main result in this section is
\begin{equation} \label{eq:mainboundR0dKL}
d^2_\text{LeCam}(p,q) \leq   d_{TV}(p,q)  \tanh \frac{1}{2 d_{TV}(p,q)}  \frac{D(p\|q) + D(q\|p)}{2}.
\end{equation}
Given that the r.h.s. of \eqref{eq:mainboundR0dKL} is an increasing function of $d_{TV}(p,q)$, we can relax it by setting $d_{TV}(p,q)$ to its maximum value (one), and obtain a novel bound in terms of symmetric Kullback-Leibler divergence alone,
\begin{equation} \label{eq:lecamleqtanh}
d^2_\text{LeCam}(p,q) \leq  \tanh \frac{1}{2}  \frac{D(p\|q) + D(q\|p)}{2}
\end{equation}
and thus, via \eqref{eq:maxRsym}, the following bound on the response intensity,
\begin{equation} \label{eq:Rpqsym-KL}
\max_f R_{p,q}(f) \leq  \frac{e^{\frac{D(p\|q) + D(q\|p)}{2}}-1}{2}.
\end{equation}
This is the second main result of this section, which we call the \emph{Symmetric Response Intensity Relation}.
Let us mention three immediate applications of these relations recovering previously known results. 

First, taking $\Omega$ as a set of trajectories for a stochastic system, $p$ the probability distribution over trajectories and $q$ the probability distribution over time-reversed trajectories, we obtain that $D(p\|q)=D(q\|p)$.
We thus see that  the Symmetric Response Intensity Relation \eqref{eq:Rpqsym-KL} contains the various TURs derived in \cite{hasegawa19,proesmans2019hysteretic,potts2019thermodynamic},
\begin{align}
\frac{\mean{f}_p^2}{\text{Var}_p(f)} \leq \frac{e^{D(p\|q)}-1}{2}
\end{align}
The added generality is that the observables here need not be current-like observables, i.e. antisymmetric under time reversal. As pointed out in \cite{falasco2020unifying}, these results hold more generally for any involution (order-two symmetry) over any probability space, and it turns out that this generality can be leveraged to provide another proof of \eqref{eq:Rpqsym-KL}, see \ref{app:equi}. 

Second, by combining the lower bound and upper bounds on Le Cam's distance, we retrieve a useful bound on the Jensen-Shannon distance in terms of Kullback-Leibler divergences, see \ref{app:JS}.

Third, an interesting observable to consider is $f=\text{sign} (p-q)$. Then we have $\langle f \rangle_{p} - \langle f \rangle_{q} =  2 d_{TV}(p,q)$ and  $\langle |f|^2 \rangle_{p}=\langle |f|^2 \rangle_{q}=1$, implying $R^{0}_{p,q}(f)= 4 d^2_{TV}(p,q)$.  
Using this in \eqref{eq:R0LeCam} (which yields $d_\text{TV} \leq d_\text{LeCam}$) and \eqref{eq:mainboundR0dKL} allows us to find the following relation between total variation distance and Kullback-Leibler divergences,
\begin{align}
d_{TV}(p,q) \leq  \tanh \frac{1}{2 d_{TV}(p,q)}  \frac{D(p\|q) + D(q\|p)}{2},
\end{align}
or
\begin{equation}\label{eq:d-vs-KL-atanh}
2 d_{TV}(p,q) \,\, \text{atanh} \,\,  d_{TV}(p,q) \leq \frac{D(p \| q)+D(q\| p)}{2}.
\end{equation}
Equivalently, as $\text{atanh} x = \frac{1}{2} \ln \frac{1+x}{1-x}$, we obtain

\begin{equation}\label{eq:d-vs-KL-ln}
d_{TV}(p,q) \ln \frac{1+d_{TV}(p,q)}{1-d_{TV}(p,q)}  \leq \frac{D(p\| q)+D(q\| p)}{2}.
\end{equation}

In fact this bound is known to be tight, since equality is achieved for binary distributions $p=(\epsilon,  1-\epsilon)$ and $q=( 1-\epsilon, \epsilon)$, for any $0<\epsilon<1$ \cite{gilardoni2006minimum}. This bound is the key ingredient for estimating the thermodynamic cost of a bit switch in the next section.

\section{Application I: trade-off between reliability and thermodynamic cost  for fast computation and other switched systems}
\label{sec:bit}

\subsection{A model for one-bit memories}

We propose the following model for a one-bit memory. We set up a time-varying Markov process over a state space $\Omega \ni \omega$ with two possible rates matrices $L_p$ and $L_q$. When the rate matrix is $L_p$, then the Markov chain converges to the (unique) stationary distribution $p$, thereby encoding a ``zero'', and converges to $q$ when the rate matrix is $L_q$, thereby encoding ``one''.  An external device switches the rate matrix from $L_p$ to $L_q$ (thus ``writing a one'' in the memory) or from $L_q$ to $L_p$ (thus ``writing a zero''). Another device reading the memory must in fact solve an inference, or decoding problem: from the value of an observable on $\Omega$ it must decide whether the Markov chain is more likely to encode a ``zero'' (i.e. obey probability distribution $p$) or a ``one'' (i.e. obey probability distribution $q$).  If $p$ and $q$ happen to have disjoint supports in $\Omega$ then the reading is errorless but usually $p$ and $q$ overlap, leaving the possibility of a wrong conclusion. We assume that the switch from $L_p$ rates to $L_q$ rates is instantaneous, or (more realistically) much faster than any other timescale of the Markov chain. Moreover, a reading, or a new writing operation occurs after the Markov process has converged to its stationary distribution (a reasonable assumption in practice).
We defer to the end of the section a more concrete description of the specific physical implementations to which this theory applies.

\subsection{Entropy production and error probability associated to a bit switch}

In this context, $\Delta \sigma_\text{switch} = k_B \frac{D(p\|q)+D(q\|p)}{2}$ has a thermodynamic interpretation: it is the non-adiabatic (also called Hatano-Sasa) entropy production associated to the relaxation process occurring in the transition between a ``zero'' and a ``one'', averaged over the two transitions. To show this we start from the splitting of the total entropy production rate $\dot \sigma_\text{total}= \dot \sigma_{na} + \dot \sigma_{hk}$ into the non-adiabatic ($\dot \sigma_{na} \geq 0$) and the adiabatic, or housekeeping ($\dot \sigma_{hk} \geq 0$) contributions, the latter being zero only for detailed balanced dynamics. The non-adiabatic entropy production rate is defined as
\begin{align}
\dot \sigma_{na} = -k_B \sum_\omega \dot p(t) \ln \frac{p(t) }{p_\text{st}(t)} = k_B \dot H(p(t)) + k_B\sum_\omega \dot p(t) \ln p_\text{st}(t),
\end{align}
where $H(p(t))=-\sum_{\omega \in  \Omega}  p(t) \ln p(t)$ is the Shannon entropy of $p(\omega, t)$ the time-dependent solution of the master equation with rates $L(t)$ switching between $L_p$ and $L_q$. The stationary solution $p_\text{st}(t)$ of the master equation is $p$ (resp. $q$) when $L(t)=L_p$ (resp. $L(t)=L_q$). Thus, the non-adiabatic entropy production associated to the transition from $p$ to $q$ is
\begin{align}
\sigma_{na}(p \to q) =k_B [ H(q)-H(p) + \sum_{\omega} (q-p) \ln q ].
\end{align}
The non-adiabatic entropy production averaged over two consecutive switches reads
\begin{align}
 \Delta \sigma_\text{switch}  = \frac {k_\text{B}}{2} (\sigma_{na}(p \to q) + \sigma_{na}(q \to p) )= k_{B} \frac{D(p\|q)+D(q\|p)}{2} ,
\end{align}
where the opposite Shannon entropy variations cancel each other.
This derivation holds even if $p$ and $q$ are non-equilibrium stationary distributions: in that case
the term above counts the entropy production associated to the relaxation excluding the housekeeping (yet time-varing) entropy production needed to sustain the memory (i.e. to maintain the violation of detailed balance) and the instantaneous work to manipulate it (i.e. switch from $L_p$ to $L_q$ and vice versa).
It is also true if $p$ and $q$ are quasi-stationary instead of truly stationary, i.e. stay approximately constant over the time scale in which reading and writing operations occur, as is the case for example in bistable (`flip-flop') memories.
In case where $p$ and $q$ are equilibria at temperature $T$, $\Delta \sigma_\text{switch} T$ is the heat dissipated into the relaxation processes of the bit switching. This is the case in some recent experimental setups \cite{berut2012,Bechhoefer14}. Conversely, $\Delta \sigma_\text{switch}$  is not directly relatable to the dissipated work in typical technological implementations in which the housekeeping entropy is non-zero (e.g. CMOS inverters or bistable memories \cite{freitas2020stochastic,freitas2021reliability}). In case of isothermal transformations, $\Delta \sigma_\text{switch} T$ is only a lower bound on the dissipation over the cycle, which is $\Delta \sigma_\text{total} T$.

The total variation distance $d_{TV}(p,q)$ also has a relevant interpretation. 
Suppose we observe the full state $\omega$ of the system (e.g. the charge in all capacitances) at a time instant $t$ and we try to infer whether the system is in a ``zero'' or a ``one'', i.e., follows distribution $p$ or $q$. If a ``zero'' or a ``one'' are a priori equally likely, then the optimal inference is the maximum likelihood estimator: if $p(\omega) > q(\omega)$ then we infer a ``zero'', otherwise we infer a ``one''. Thus the observable $f=\text{sign} (p-q)$ is the optimal ``reading'' of the bit. In this context, $\epsilon=\frac{1-d_{TV}(p,q)}{2}$ is precisely the probability of error, i.e. of reading a ``zero'' while the system obeys distribution $q$ or a ``one'' when it obeys $p$. Hence, the Response Intensity Relation for the sign observable \eqref{eq:d-vs-KL-ln}  can be rewritten as
\begin{equation}\label{eq:eps-sigma}
k_B (1-2 \epsilon) \ln(\epsilon^{-1}-1) \leq \Delta \sigma_\text{switch} \leq \Delta \sigma_\text{total},
\end{equation}
where in the last equality we used that the total entropy production $\Delta \sigma_\text{total}$ of the switch is obtained by adding the non-negative housekeeping entropy production.
Equation \eqref{eq:eps-sigma} expresses a relation between error probability and thermodynamic cost of the computation. We remind that this relation---the main result of this section---is deduced under the assumption of a Markov process undergoing an abrupt (instant) switch between the two dynamics.

\subsection{Reliable vs noisy computations}\label{sec:computation}

In the limit of reliable computations, $\epsilon \ll 1$, \eqref{eq:eps-sigma} simplifies to $ k_B
\ln \epsilon^{-1} \leq \Delta \sigma_\text{switch}$, or equivalently 
\begin{align}\label{eq:error_reliable}
\epsilon \geq e^{-\Delta \sigma_\text{switch}/k_B}.
\end{align}
A similar relation has been obtained under particular technological assumptions fitting  the current technological paradigm (e.g., circuit with linear capacitances operating under Gaussian thermal noise) \cite{kish2004moore}. 

In the opposite noisy computation limit, $d_{TV} = 1-2\epsilon \ll  1$ (i.e. $\epsilon \approx 1/2$), \eqref{eq:eps-sigma} simplifies to
\begin{align}
2 k_B d^2_{TV} \leq \Delta \sigma_\text{switch}.
\end{align}
Note that in this limit, the output has very low information in the sense of Shannon, i.e. the mutual information between writing (alternating between $L_p$ or $L_q$ with probability $1/2$) and the reading is very low (with error probability close to $1/2$). As a matter of fact, the mutual information is equal (in nats) to 
\begin{align}
I=\ln 2 - [\epsilon \ln \epsilon^{-1} +  (1-\epsilon) \ln (1-\epsilon)^{-1})] \approx  d^2_{TV}/2,
\end{align}
where the first term is the Shannon entropy of the writing process and the second term in square brackets is the conditional entropy of the writing given output of the reading process. Therefore, the cost of switching a bit in this context is 
\begin{equation}\label{eq:I-sigma}
4 k_B I \leq \Delta \sigma_\text{switch} \leq  \Delta \sigma_\text{total}.
\end{equation}
To speak in terms of energy, the cost of switching a bit abruptly in an environment of temperature $T$ is at least $4k_BT$ per nat, or $2.8 k_BT$ per Shannon bit, independently of the (low) reliability. This is in contrast with the low noise limit \eqref{eq:error_reliable}, where the cost per bit switch grows unboundedly as we approach perfect reliability. This supports quantitively the common knowledge that noisy circuits may be more energetically efficient than low-noise  circuits. Of course, managing noisy information may require carefully designed noise-tolerant algorithms \cite{shanbhag2002reliable,alaghi2017promise,freitas2020stochastic}.

\subsection{Discussion on the energy-error-speed trade-off}

To make things more concrete, we consider the example of the NOT gate  (also known as the inverter), an elementary electronic circuit ubiquitous in digital computing devices, as it is a building block for more complex circuits. This circuit is controlled by an external voltage $u(t)$ which takes the value $u_{\min}$ or  $u_{\max}$. This voltage determines the random dynamics of the charges flowing into the capacitances of the inverter. When stationarity is reached, an output capacitance has a \emph{mean} voltage close to $u_{\max}$ (encoding a ``one'') when the input voltage is $u_{\min}$ and conversely. The reading of the bit is in practice a thresholding of the output capacitance voltage: if it takes a value larger than $(u_{\min}+u_{\max})/2$ we consider it encodes a ``one'', otherwise a ``zero''. Due to random fluctuations however, an erroneous reading may occur.  The technological details do not matter here: it suffices to know that the dynamics of such a circuit is well modeled indeed by a Markov process. We refer the reader to \cite{freitas2020stochastic,freitas2021reliability} for explicit and realistic constructions of Markov chains for inverters and more complex memories. 

We made the assumption of an abrupt, instant switching between two rate matrices. In the practice of a digital circuit such as an inverter, this would amount to an instant switch between voltages $u_{\min}$ and $u_{\max}$. Since these voltages are themselves the output of another circuit, they do not evolve instantly but with a certain time constant. Taking these time scales into account would need to correct the analysis above. Nevertheless, these time constants tend to be made as short as possible in current digital technologies, and the abrupt switching can be seen as an ideal limiting case.

Let us remind that the actual reading of the bit in practice is not always given by the maximum-likelihood binary  observable $\text{sign}(p-q)$ as we have assumed here, but for instance by $\text{sign}(f)$, where $f$ is a physical observable such as the voltage on one specific capacitance. If the two differ, the reading process may be suboptimal, resulting in an extra error probability. From Chebyshev-Cantelli's inequality alone, assuming $f$ centered and $\text{Var}_p f=\text{Var}_q f$ for simplicity, we know that an inference based on $\text{sign}(f)$ will have an error $\epsilon \leq (1+ R_{p,q}(f)/4)^{-1}$, which is a very slow decrease. If more is known about the distribution of $f$, for instance if it is Gaussian, then we may derive an exponentially fast decrease of the error with  $R_{p,q}(f)$, as indicated from Chernov's bound for instance, or bounds in \cite{duembgen2010bounding}.

Remember that the main result of this section, \eqref{eq:eps-sigma}, is derived from \eqref{eq:lecamleqtanh} applied to the observable $\text{sign}(p-q)$ (which is the outcome of the bit reading process). If our goal is to obtain a lower bound on $\sigma_\text{switch}$ that is as tight as possible, one should rather use the response intensity of the maximum-response-intensity observable $m=2(p-q)/(p+q)$ or an observable close to it. In the high reliability limit ($\epsilon \ll 1$), the distributions $p$ and $q$ have a small overlap, and the binary observable $\text{sign}(p-q)$ is always close to $m$ indeed. In the low accuracy limit however, the two might differ substantially, and the response of other observables may offer a better lower bound on entropy production. Consequently, the bound \eqref{eq:I-sigma} may be loose in these cases. 

Importantly, note that the bound \eqref{eq:I-sigma} is unrelated to the similar-looking Landauer's bound. Landauer's bound concerns the \emph{erasure} of a bit, and is relevant universally. We know, in agreement with Landauer's bound, that switching a bit from zero to one and conversely can in principle be done at zero dissipation, with underdamped time-varying circuits or with infinitely slow overdamped circuits. We place ourselves at another end of the spectrum, with abrupt (as fast as possible) switching in a circuit undergoing a relaxation to stationarity. A full theory encompassing the energy-accuracy trade-off in all the spectrum of speeds is beyond the scope of this paper. Our results here are, as far as we know, among the first towards a trade-off for general classes of systems at non-zero speeds. See however \cite{proesmans2020landauer} for energy-optimal erasure in finite time on the real line.

Finally, even though we discussed the case of a bit switch in this section, note that dynamical systems switching abruptly between two or more regimes are common in physics and engineering beyond the realm of computation, and the results discussed here can in principle be adapted for those (see \ref{app:switch}). More abstractly, every inference problem involving a binary decision can benefit from the same analysis. For example, in \cite{rol15} it was considered the problem of deciding whether a given trajectory of a stationary Markov chain is unfolded forward or backward in time. Eq. (9) in \cite{rol15} coincides with our \eqref{eq:eps-sigma} with $\Delta \sigma_\text{switch}$ being the total entropy production of the process. 
Adapting our derivation to this context (see \ref{app:arrow}), we can indeed extend the validity of the result of \cite{rol15}, showing that it  holds under very general circumstances, e.g. independently of the choice of a certain stopping criterion and for non-Markovian dynamics .

 \section{Application II: Classical Speed Limits}

 \subsection{What is a Classical Speed Limit?}
 
 Classical Speed Limits (SLs) are lower bounds on the time needed for a driven system to reach a target state from an initial state, in terms of thermodynamic or kinetic quantities. The main such result \cite{shi18, vo20} states that if a non-autonomous finite-state Markov chain starts from state probability $p(0)=p$ and finishes in state probability $p(\tau)=q$, then the time $\tau$ to do so is lower bounded as follows:
\begin{equation}\label{eq:standardCSL}
\tau  \geq \frac{2d^2_{TV}(p,q)}{\Delta \sigma \overline{A}}.
\end{equation}
In this inequality, $\Delta \sigma$ is the (non-adiabatic, or Hatano-Sasa) entropy production along the trajectory that is associated with the relaxation to the instantaneous stationary state. The latter is defined at each time $t$ as the probability distribution that would be reached by the system if the driving stopped, i.e. the transition rate became time-independent for all times larger than $t$. The activity $\overline{A}$ is the time-average activity rate, i.e. expected number of jumps per time unit.
In \eqref{eq:standardCSL}, a thermodynamic term (non-adiabatic entropy production) quantifies how far we stand from the stationary distribution, and a kinetic term (activity) quantifies how fast the system evolves.

In this section we adopt the view that a classical SL is any inequality that can be written in the form

\begin{equation}\label{eq:templateCSL}
\tau \geq h(\text{Dist}, \text{Therm}, \text{Kin})
\end{equation}
where $\text{Dist}$ is the distance between $p$ and $q$, for any notion of distance between probability distributions, $\text{Therm}$ is a thermodynamic quantity capturing how far $p(t)$ is from equilibrium or stationarity, and  $\text{Kin}$ is a kinetic quantity which measures the `fastness' of the dynamics.  If the time $\tau$ and the distance are very small (infinitesimal), we can speak of \emph{local classical SL}.
We leverage the concepts encountered above, such as the maximum response intensity between $p$ and $q$, to derive new classical SLs involving various statistical distances, thermodynamic and kinetic quantities of interest. 

\subsection{Bounding Le Cam's distance with an integral along the trajectory}
Le Cam's distance $d_\text{LeCam}(p,q)$, via \eqref{eq:maxRsym}, captures the maximum response intensity between $p$ and $q$. The key observation is that the triangle inequality allows to write

\begin{align}
d_\text{LeCam}(p,q)  \leq \sum_{k=0}^{K-1} d_\text{LeCam}(\,p(k\Delta t),p(\,(k+1)\Delta t\,)\,)
\end{align}
for $\tau = K \Delta t$, where $\Delta t$ can then be taken arbitrarily small, and $K$ very large.
In the limit of very small $\Delta t$, Le Cam's distance essentially coincides with Fisher information metric:

\begin{align}
d^2_\text{LeCam}(p,p + \Delta p) = \frac{1}{4}\sum_{\omega} \frac{\Delta p^2}{p} + \mathcal{O}(\Delta p^3).
\end{align}

From the triangle inequality and the local behaviour of the distance, we find:

\begin{equation} \label{eq:LeCampath}
d_\text{LeCam} (p,q) \leq \frac{1}{2} \int_0^{\tau} \bigg\| \frac{\dot{p}}{\sqrt{p}}\bigg \|_2 dt 
\end{equation}
where $\|.\|_2$ is the usual Euclidean norm (square root of sum of squares).
To go further, we need to  compute and bound $\dot{p}$, as we do in next sections.

\subsection{Forward and backward master equation}
For Markov chains the evolution of $p(t)$ is dictated by a transition rate matrix $L(t)$, to which an instantaneous stationary distribution $p_\text{st}(t)$ is associated.
We thus have the master equation 
\begin{align}
\dot{p}(t) = p(t) L(t),
\end{align}
as commonly written in Markov chain literature notation, i.e. with the state probabilities as a row vector, and the matrix elements $L_{\omega \omega'}$ ($\omega \neq \omega'$) denoting the transition rate from state $\omega$ to state $\omega'$. The diagonal entry is $L_{\omega\omega}= - \sum_{\omega' \neq \omega} L_{\omega\omega'}$, so that each row of the matrix $L$ sums to zero. 
We assume for simplicity that at each time instant, $L(t)$  admits a unique stationary probability distribution  $p_\text{st}(t)$, defined as the left eigenvector satisfying $p_\text{st}(t) L(t)=0$.


As it will be useful later, we also write the master equation backward in time,
\begin{align}
-\dot{p}(t) = p(t) L^*(t)
\end{align}
where the transition rate matrix is $L^*=P^{-1} L^T P - P^{-1} \dot{P}$, and $P=\text{diag}(p)$ is the diagonal matrix created from $p$. This is proved as follows: the off-diagonal entries of $PL$ and $(PL^*)^T$ coincide, because the entry $\omega \omega'$ of $PL$ represents the probability flow from state $\omega$ to state $\omega'$, and the entry $\omega' \omega$ of $PL^*$ represents the probability flow from state $\omega'$ to state $\omega$ in the backward dynamics, which is of course the same. The diagonal entries of $L^*$ must be such that each row of $L^*$ sums to zero. 
The diagonal entries of $L$ represent the \emph{escape rates} from each state, while the diagonal entries of $L^*$ represent the escape rates in the backward dynamics, thus the \emph{entrance rates} in the forward dynamics \cite{falasco15inflow}.


\subsection{Local Speed Limit}

In order to estimate \eqref{eq:LeCampath}, we want to find an upper bound on $\|\dot{p} P^{-1/2} \|_2$, knowing that $\dot{p}=pL$. This will lead to  what we can call a local classical SL.

Knowing that $p_\text{st}L = 0$ (expressing that $p_\text{st}$ is a stationary state), we can write $\dot{p} = (p-\alpha p_\text{st}) L$, for any scalar coefficient $\alpha$ of our choice. We can also write:

\begin{align}
\dot{p} P^{-1/2}&= (p-\alpha p_\text{st}) L P^{-1/2}\\
		&= (\sqrt{p} - \alpha  p_\text{st} P^{-1/2}) P^{1/2} L P^{-1/2} \label{eq_pdotP}
\end{align}

It is useful to introduce the two-norm of a square matrix $A$, defined as the maximum ratio $\frac{\|x A\|_2}{\|x\|_2}$ over all non-zero (row) vectors $x$. It turns out to be equal to the maximum singular vector of $A$, i.e. the square root of the spectral radius $\rho$ (eigenvalue of largest magnitude) of the symmetric matrix $AA^T$. This is denoted
\begin{align}
\|A\|_2 = \sqrt{\rho(AA^T)}=\sqrt{\rho(SAA^TS^{-1})},
\end{align}
for any invertible square matrix $S$ (encoding a linear change of coordinates, which does not affect eigenvalues).
Thus we obtain from \eqref{eq_pdotP}:
\begin{align} \label{eq:prelocalCSL}
\|\dot{p} P^{-1/2} \|_2 \leq \|\sqrt{p} - \alpha  p_\text{st} P^{-1/2}\|_2 \| P^{1/2} L P^{-1/2}\|_2.
\end{align}
First, we estimate $\| P^{1/2} L P^{-1/2}\|_2$ as
\begin{align}
	\| P^{1/2} L P^{-1/2}\|_2 &= \sqrt{\rho( P^{1/2} L P^{-1/2}  P^{-1/2} L^T P^{1/2})}\\&= \sqrt{\rho( L P^{-1}  L^T P)} \\
	&\leq \sqrt{\| L\|_{\infty} \|P^{-1}  L^T P\|_{\infty}}.
\end{align}
Here we have used the matrix norm $\|A\|_{\infty}$, which is an upper bound on the spectral  radius ($\rho(A) \leq \|A\|_{\infty}$) and submultiplicative ($\|AB\|_{\infty} \leq \|A\|_{\infty}\|B\|_{\infty}$). It is defined as the maximum one-norm of a row of $A$, or equivalently as $\max \frac{\|xA\|_1}{\|x\|_1}$. Thus $\| L\|_{\infty}$ is twice the maximum escape rate from each state, which we denote $A_\text{max,escape}=\max_\omega L_{\omega\omega}$. The one-norm of $P^{-1}  L^T P$ is obtained by noticing that the off-diagonal entries of this matrix are the same as $L^*$ (the matrix of backward rates), while the diagonal entries are the same as $L$. Thus the one-norm of $P^{-1}  L^T P$, denoted $A_\text{max,entrance}$, is the largest sum of escape rate and entrance rate (which is the escape rate in the backward random walk) on a state. 
Therefore, we arrive at
\begin{align}
\| P^{1/2} L P^{-1/2}\|_2 &\leq \sqrt{\| L\|_{\infty} \|P^{-1}  L^T P\|_{\infty}}\\
&\leq \sqrt{2A_\text{max,escape} (A_\text{max,exit}+A_\text{max,entrance})}\\
&\leq 2 \max (A_\text{max,escape},  A_\text{max,entrance}) \\
&=2	A_\text{max},	
\end{align} 
where $A_\text{max}= \max (A_\text{max,escape},  A_\text{max,entrance})$ is the \emph{maximum activity rate}.
Second, we estimate $\|\sqrt{p} - \alpha  p_\text{st} P^{-1/2}\|_2$:
\begin{align}
\|\sqrt{p} - \alpha  p_\text{st} P^{-1/2}\|_2^2 &= \sum_{\omega}\left[ p(\omega) -2 \alpha  p_\text{st}(\omega)   + \alpha^2 p^2_\text{st}(\omega) / p(\omega) \right] \\
&= 1-2\alpha + \alpha^2 \sum_\omega p^2_\text{st}(\omega) / p(\omega) \\&= 1-2\alpha +  (1+\chi^2(p_\text{st} \| p)) \alpha^2 \label{eq:alphachi2}
\end{align}
Choosing $\alpha= 1/(1+\chi^2(p_\text{st} \| p))$, which minimizes \eqref{eq:alphachi2}, we obtain:
\begin{align}
\|\sqrt{p} - \alpha  p_\text{st} P^{-1/2}\|_2^2 &= \frac{\chi^2(p_\text{st} \| p)}{1+\chi^2(p_\text{st} \| p)} \\
&\leq \min(1, \chi^2(p_\text{st} \| p))
\end{align}
With these estimates, our bound \eqref{eq:prelocalCSL} turns into the local classical SL  
\begin{equation}\label{eq:LocalCSL}
\|\dot{p} /  \sqrt{p} \|_2 \leq 2 \sqrt{ \frac{\chi^2(p_\text{st} \| p)}{1+\chi^2(p_\text{st} \| p)}} A_\text{max}. 
\end{equation}
 This is indeed a classical SL in the form \eqref{eq:templateCSL} as an upper bound on the speed $\dot p$ is also a lower bound on the (short) time $\Delta t$ needed to achieve a certain (short) displacement $\Delta p = \dot p \Delta t+ \mathcal{O}(\Delta t^2)$. Referring to \eqref{eq:templateCSL}, our thermodynamic quantity is $\chi^2(p_\text{st} \| p)$, which quantifies, as we have seen, the maximum response intensity to the change from $p$ to stationarity $p_\text{st}$, thus capturing in a meaningful way whether $p$ is far from stationarity. Our kinetic quantity is $A_\text{max}$, which bounds the entrance or escape at any state. In other words, $1/A_\text{max}$ can be interpreted as the fastest time scale at any state in any direction of time.   

\subsection{The classical SL for Le Cam's distance}

Integrating the local bound \eqref{eq:LocalCSL} into \eqref{eq:LeCampath}, we obtain a SL for the time interval $[0,\tau]$ between $p=p(0)$ and $q=p(\tau)$:
\begin{equation}\label{eq:LeCamintegral}
d_\text{LeCam}(p,q) \leq \int_0^{\tau} \sqrt{ \frac{\chi^2(p_\text{st} \| p)}{1+\chi^2(p_\text{st} \| p)}}  A_\text{max} dt
\end{equation}
In order to obtain separate thermodynamic and kinetic factors, we can for instance bound the integral as follows:
\begin{equation}\label{eq:holder0infty}
\left|\int_0^{\tau} fg \, dt \right | \leq \sup_t |f|  \,\,\int_0^{\tau} |g|\,  dt.
\end{equation}
In this way we get a novel classical SL:

\begin{equation}\label{eq:LeCamspeedlimit}
d_\text{LeCam}(p,q) \leq \sup_{0 \leq t \leq \tau} \sqrt{ \frac{\chi^2(p_\text{st} \| p)}{1+\chi^2(p_\text{st} \| p)}} \,\, \overline{A}_\text{max} \,\, \tau
\end{equation}
where $\overline{A}_\text{max}= \tau^{-1}\int_0^{\tau} A_\text{max}(t) dt$ is the time-average of the maximum activity. This evidently provides a lower bound on $\tau$, in terms of total distance, thermodynamic and kinetic factors, following the template \eqref{eq:templateCSL}.
Note that since the thermodynamic term is bounded by $1$, we have the purely kinetic SL:
\begin{equation}\label{eq:LeCamkinetic}
d_\text{LeCam}(p,q) \leq  \overline{A}_\text{max} \tau
\end{equation}

This bound is acceptable far from stationarity, when $\chi^2(p_\text{st} \| p) \gg 1$ for some part at least of the trajectory. Nevertheless, the thermodynamic factor is useful when driving close to stationarity, when 
$\chi^2(p_\text{st} \| p) \ll 1$ (low asymmetric response intensity), yielding a much tighter bound than \eqref{eq:LeCamkinetic}.

\subsection{A classical SL with the Total Variation Distance}

We now apply the same methodology as for Le Cam's distance to the Total Variation distance.
Locally, we find:
\begin{align}
d_{TV}(p(t), p(t+\Delta t))=\frac{1}{2} \|\Delta p\|_1 = \frac{1}{2}  \|\dot{p} \|_1 \Delta t + \mathcal{O}(\Delta t^2)
\end{align}
Thus, from the triangle inequality we obtain
$$
d_{TV}(p,q)\leq \frac{1}{2} \int_0^{\tau} \|\dot{p}(t)\|_1 dt ,
$$
in which the integrand in the r.h.s. can be estimated as
\begin{align}
\frac{1}{2}\|\dot{p}(t)\|_1 &= \frac{1}{2} \|pL\|_1 \\ &= \frac{1}{2} \|(p-p_\text{st}) L\|_1 \\&\leq   \frac{1}{2} \|(p-p_\text{st}) \|_1 \|L\|_\infty \\&\leq \frac{1}{2} \|(p-p_\text{st}) \|_1 \, 2 \max_\omega |L_{\omega\omega}| \\&= 2 d_\text{TV}(p,p_\text{st})  A_\text{max,exit},
\end{align}
yielding
\begin{equation}\label{eq:CSLTVintegral}
d_{TV}(p,q)\leq  2 \int_0^{\tau} d_\text{TV}(p,p_\text{st})  A_\text{max,escape} dt.
\end{equation}
Using \eqref{eq:holder0infty}, the latter can be further bounded as
\begin{equation}\label{eq:CSLTVTV}
d_{TV}(p,q)\leq  2 \sup_{0 \leq t \leq \tau} d_\text{TV}(p,p_\text{st})\,\, \overline{A}_\text{max,escape} \,\, \tau ,
\end{equation}
which constitutes an example of classical SL involving the total variation distance both as the distance between initial and final state, and as a thermodynamic measure of distance to stationarity. It can be related to available free energy $D(p\|p_\text{st})$ through Pinsker's inequality
\begin{align}
  d_\text{TV}(p,p_\text{st}) \leq \sqrt{\frac{1}{2}D(p\|p_\text{st})},
\end{align}
thus giving the classical SL
\begin{equation}\label{eq:CSLTVTV_pinsker}
d_{TV}(p,q)\leq  \sup_{0 \leq t \leq \tau} \sqrt{2 D(p\|p_\text{st})} \,\, \overline{A}_\text{max,escape} \tau 
\end{equation}
It resembles the original classical SL obtained in \cite{shi18, vo20}, although they are not directly comparable. The main conceptual difference is that here our thermodynamic factor is the total available free energy (in case of detailed balanced dynamics), rather than its dissipation rate.

Note that this can be improved by using other bounds on total variation distance than Pinsker's, for instance Bretagnolle-Huber's bound \cite{bretagnolle1979estimation}
\begin{align}
d_\text{TV}(p,p_\text{st}) \leq \sqrt{1 - e^{-D(p\|p_\text{st})}},
\end{align}
which is worse than Pinsker's close to stationarity but better far from stationarity. We thus find another classical SL:
\begin{equation}\label{eq:CSLTVTV_BH}
d_{TV}(p,q)\leq 2  \sup_{0 \leq t \leq \tau} \sqrt{1 - e^{-D(p\|p_\text{st})}} \,\, \overline{A}_\text{max,escape} \tau 
\end{equation}
From this  we have again a purely kinetic speed limit:
\begin{equation}\label{eq:CSLTVTV_kin}
d_{TV}(p,q)\leq  2 \overline{A}_\text{max,escape} \tau 
\end{equation}
This speed limit is useful if $p$ is far from stationarity (total variation distance close to one), but extremely loose when driving the distribution in a quasi-stationary fashion. An analogous fact is true for precision: kinetic uncertainty relations \cite{dit18} are tighter that thermodynamic ones for large entropy production.

Note that one can easily obtain many variants of the speed limits, following the same methodology with different bounds or distances. A few ones are highlighted in \ref{app:csl}.

\section{Application III: Simplicity implies robustness, complexity implies fragility}

Given a reference distribution $q$ on a space $\Omega$, and a perturbed distribution $p$, the maximum of (asymmetric) response intensity over all observables on $\Omega$ is $\chi^2(p\|q)=\sum_{\omega}p^2/q -1$, as derived in section \ref{sec:sym_resp}.

\subsection{In trajectory space: Langevin dynamics}
First, we focus on $\Omega$ being the space of trajectories of a stochastic system and we rewrite the probability density $p(\omega)= e^{\mathcal{A}_p(\omega)}$ (resp. $q(\omega)= e^{\mathcal{A}_q(\omega)}$) as the exponential of an action functional. To be concrete, we consider the case of diffusive dynamics given (in the unperturbed case $q$) by the Langevin equation
\begin{align}\label{eq:langevin}
\dot x(t)= \mu F(x(t)) + \sqrt{2D} \xi(t).
\end{align}
for the vector $x \in \mathbb{R}^{dN}$ describing, e.g, $N$ particles in $d$ dimensions.
Equation \eqref{eq:langevin} features a delta-correlated Gaussian noise $\xi$, a positive definite symmetric diffusion matrix $D$ and mobility matrix $\mu$ (not necessarily connected by the Einstein relation $D= k_B T \mu$).
It corresponds to the action 
\begin{align}
\mathcal{A}_q(\omega)= -\frac 1 4 \int_0^\tau dt (\dot x(t)-  \mu F(x(t))) D^{-1} (\dot x(t)-  \mu F(x(t)))^T
\end{align}
for the trajectories $\omega=\{x(t): t \in [0,\tau] \}$. If $p$ and $q$ are absolutely continuous, i.e. the ratio $p/q$ (more rigorously, the Radon-Nikodym derivative) exists and is finite, the computation of the maximal response reduces to an exponential average over unperturbed  trajectories
\begin{align}\label{eq:chi_langevin}
\chi^2(p\|q)= \mean{e^{2(\mathcal{A}_p-\mathcal{A}_q )}}_q-1,
\end{align}
that might be much less involved than an average w.r.t. the perturbed $p$ ($q$ can refer to an equilibrium or non-interacting system, for instance).
Perturbations that satisfy \eqref{eq:chi_langevin} are, for example, changes of the drift, such as $F(x) \to F(x) + g(x)$. Instead, a perturbed probability density $p$ generated by a change in the matrix $D$ is not absolutely continuous w.r.t. $q$ (and response in trajectory space is known to require explicit regularization procedures \cite{falasco2016nonequilibrium,falasco2016temperature}). Restricting us to a change in the external forces $F(x) \to F(x) + g(x)$, we obtain 
\begin{align}\label{eq:chi_langevin2}
\chi^2(p\|q)= \mean{e^{ \int_0^\tau dt [\mu g D^{-1} (\dot x - \mu F) -\frac 1 2 \mu g   D^{-1}  (\mu g)^T] }}_q-1.
\end{align}
In general \eqref{eq:chi_langevin2} is hard to compute as it involves averaging a functional of the trajectories and thus requires knowledge of the time propagator. Possibly, a saddle point calculation can be performed in some large-deviation regimes, e.g., long time $\tau \to \infty$, weak noise $D \to 0$ (entry-wise), and in some proper thermodynamic limit involving $N \to \infty$.
For small perturbations, we retrieve Dechant-Sasa's bound on the linear response  (see \eqref{eq:linear_resp}),
\begin{align}
\max_f R_{p\|q}(f) = \frac 1 2  \int_0^\tau dt  \mean{\mu g   D^{-1}  (\mu g)^T}_q.
\end{align}
Here we point out how the complexity of the system (spread in space and range of interactions) potentially leads to large responses in paradigmatic models of energy and mass transport. 
To show that we consider as the unperturbed systems independent particles in a stationary state, subject to a single-particle force $F(x)=(F_1(x_1),\dots, F_N(x_N))^T$ and in contact with separate thermal baths at temperature $T_i$.
Hence, we set $\mu$ equal to the identity matrix (choosing appropriate time units) and $D$ diagonal with elements $D_{ii}= k_B T_i \mu_i$. The perturbation $g(x)=(g_1(x), \dots, g_N(x))$ may consist of an arbitrary interaction force coupling different particles. The maximal linear response is thus given by
\begin{align}\label{eq:max_linear_langevin}
\frac{1}{\tau}\max_f R_{p\|q}(f) =  \sum_{i=1}^N\mu_i \frac{\mean{\|g_i \|_2^2}_{q} }{2 k_B T_i}   \leq M^2 \kappa^2  \sum_{i=1}^N \frac{ \mean{\|x_i \|_2^2}_{q} }{2 k_B T_i},
\end{align}
where the second equality holds by assuming that the interactions can be modeled to leading order as a linear force of strength $\kappa$ involving $M$ neighbors, and by taking $x=(x_1, \dots, x_N)^T$ as the displacement from the particle rest positions (i.e. $\mean{x}_q=0$).
This general setting covers standard models of heat conducting lattices and loaded molecular motors (see e.g. \cite{falasco15harmonic, falasco20dissip}). In the first case, $F(x)$ is an arbitrary nonlinear potential force; in the second $F(x)$ is typically a tilted sinusoidal potential, the dynamics is restricted by spatial periodic boundary conditions within a period, and $T_i=T$ for all $i$. 
Note that for the observables of interest, such as the time-integrated heat and particle currents, the scaling by $1/\tau$ ensures that the l.h.s of \eqref{eq:max_linear_langevin} is independent of $\tau$.
In both cases the more `complex' is the system (larger spreading in space $ \mean{\|x_i \|_2^2}_{q}$) and the perturbation (larger range of interactions $M$), the larger the maximal linear response.


\subsection{In finite state spaces}
Second, we take $\Omega$ to be the state space of a stochastic system. For a finite space $\Omega$, we now consider the maximum $\chi^2(p\|q)$ over all possible perturbations $p$. The perturbation leading to highest possible response is the distribution $p$ concentrated on $\omega$ such that $q(\omega)=q_{\min}$ is minimal. Thus 

\begin{align}
\max_p \chi^2(p\|q)=\frac{1}{q_{\min}}-1
\end{align}

We take $1/q_{\min}$ as an index of complexity of the system. If $\Omega$ has cardinality $N$ then $1/q_{\min} \geq N$. Inversely, if $q$ is relatively uniform over a few states, then $1/q_{\min}$ is small. This is quantified by the following inequalities:

\begin{align}
e^{H(q)} \leq N \leq \frac{1}{q_{\min}} \leq e^{H(q)}  \frac{q_{\max}}{q_{\min}} \leq N \frac{q_{\max}}{q_{\min}},
\end{align}
where $H(q)=\sum_{\omega \in  \Omega}  q \ln \frac{1}{q}$ is the Shannon entropy, with bounds  
\begin{align}
  \frac{1}{q_{\max}} \leq  e^{H(q)} \leq  N \leq \frac{1}{q_{\min}}.
\end{align}
 
We can always write $q(\omega)$ in an exponential form $q(\omega)=e^{-\beta (E(\omega)-F)}$, where $F= \langle E \rangle - \beta^{-1} H(q)$. This is especially relevant when $q$ is an equilibrium distribution, where $E$ is an energy, $\beta^{-1}=k_BT$ and $F$ the equilibrium free energy.
In this case we can write:

\begin{align}\label{eq:qmin}
\frac{1}{q_{\min}} = e^{-\beta E_{\max}+F} \leq   e^{H(q)} e^{\beta(E_{\max}-E_{\min})}
\end{align}

Thus as a general rule we find that `simple' systems (with low $1/q_{min}$) are `robust', in that the response of any observable to any perturbation does not go beyond a few standard deviations.  Conversely, `complex' systems (e.g. with a large entropy or large energy spreading) are fragile, in that some observables are highly sensitive to some perturbations. This is reminiscent, although in another context, of the `complexity implies fragility' principle advocated in \cite{carlson2002complexity,ames2008complexity}. If we see the response as a useful signal rather than a disturbance, then this means that simple systems are unable to transmit with a high signal-to-noise ratio. 

This allows some insight on the value of simple models.  Consider that typically nonequilibrium toy models have few states, e.g. $N=3$ for a minimal Markov chain able to sustain a current (without recurring to multiple transitions between two states). A typical example is offered by the enzyme kinetics describing substrate inhibition as a Markov chain on the three different states of the enzyme (free, and bounded to one or two substrate molecules) \cite{fal19n,forastiere2020strong}.
If we mean to describe a biophysical process in this way, then typically the energy spread of the unperturbed system can be taken to be $E_{\max} - E_{\min}  \simeq k_B T$ . 
Therefore, we find $|\langle f \rangle_p - \langle f \rangle_q|  \lesssim \sqrt{3 e^1-1} \sqrt{\text{Var}_{q}} \approx 2.67 \sqrt{\text{Var}_{q}}$, namely, the response never exceeds 3 standard deviations irrespective of the perturbation strength. Thus, few-state models are by construction robust. Also, if results of an experiment were rationalized with a few-state model, they would be unable to tell whether the measured system is in or out of equilibrium. 

\section{Conclusions}

We have explored bounds on the response of observables subject to a perturbation of the underlying probability distribution, introducing the response intensity as an adimensional ratio comparing the size of the response to the size of typical fluctuations. It turns out that bounds on the response intensity can be found by the Hilbert Uncertainty Relation, a recent  generalization of TURs. These bounds allow to derive several nontrivial inequalities between the various distances or divergences that can be defined between probability distributions, some of them well-known and some of them novel. We believe that the novel speed limitis we have subsequently derived and the novel applications we have provided on switched systems (e.g. computing devices) and biophysical systems (e.g. the significance of simple models), together with other recent results of the stochastic thermodynamics literature, are only the tip of the iceberg of an emerging theory that characterizes the necessary relations between kinetics, thermodynamics and reliability of stochastic systems. 
   
 \section{Acknowledgements}
 
J-C. D. and M.E. acknowledge funding from the project INTER/FNRS/20/15074473
	``TheCirco'' on Thermodynamics of Circuits for Computation, funded by the F.R.S.-FNRS (Belgium) and
	FNR (Luxembourg).

\appendix

\section{Lower bounds on Le Cam's distance}\label{app:lower}
	On the one hand, we derive from \eqref{eq:chivsKL} lower bounds on Le Cam's distance:
	\begin{align}\label{eq:lecamgeqasymKL}
		d^2_\text{LeCam}(p,q) \geq e^{D(p\|\frac{p+q}{2})}-1 \geq D(p\|\frac{p+q}{2})
	\end{align}
	and
	\begin{align}
		d^2_\text{LeCam}(p,q) &\geq \frac{1}{2} e^{D(p\|\frac{p+q}{2})} + \frac{1}{2} e^{D(q\|\frac{p+q}{2})} - 1 \\ 
		&\geq e^{\frac{1}{2}(D(q\|\frac{p+q}{2})+D(p\|\frac{p+q}{2}))} - 1\\
		&\geq \frac{1}{2} (D(p\|\frac{p+q}{2}) + D(q\|\frac{p+q}{2}))\\
		&=d^2_{JS}(p,q)  \label{eq:JSvsLeCam}
	\end{align}
	The latest quantity, $d_{JS}(p,q)$, defined as the square root of of  $\frac{1}{2}(D(p\|\frac{p+q}{2}) + D(q\|\frac{p+q}{2}))$, is also a distance (satisfying the triangle inequality), called  the \emph{Jensen-Shannon distance}.

\section{Equivalence of symmetric response intensity bounds with the TURs for antisymmetric observables}\label{app:equi}

TURs were initially devised for $\Omega$ being the set of trajectories of a (stochastic) physical system, and $\mathcal{F}$ being the set of observables anti-symmetric under time-reversal (and possibly other constraints). In a more abstract version, regardless of any thermodynamic context, one considers an abstract $\Omega$ endowed with an involution $s:\Omega \to \Omega$ (i.e. a bijection whose square is the identity), which is not necessarily time-reversal but could be space reversal, spin reversal, etc. \cite{falasco2020unifying}.
The relations obtained as upper bounds on symmetric nonlinear response, such as \eqref{eq:Rpqsym-KL} or \eqref{eq:mainboundR0dKL},  generalize results obtained previously in this literature, for $p$ being the probability distribution on $\Omega$, and $q$ being the symmetric distribution $q(\omega)=p(s\omega)$.

Conversely, these known results offer an alternative way to rederive the nonlinear response bounds, thanks to the following trick. Two distributions $p$ and $q$ on one same space $\Omega$ can be recoded into a single distribution on the space $\Omega \oplus \Omega$ (the disjoint union of two copies of $\Omega$), with probability $p/2$ on one copy of $\Omega$ and $q/2$ on the other copy. The involution $s: \Omega \oplus \Omega \to  \Omega \oplus \Omega$ simply swaps the two copies  of $\Omega$. The observables $f$ on $\Omega$ can be recoded into antisymmetric observables $f \oplus (-f)$ on $\Omega \oplus \Omega$. Applying the various bounds derived in \cite{falasco2020unifying} on antisymmetric observables we recover the bounds \eqref{eq:mainboundR0dKL} or \eqref{eq:Rpqsym-KL}. 

In conclusion there is an equivalence between the Symmetric Response Intensity Relation \eqref{eq:Rpqsym-KL} and the TUR for antisymmetric observables with respect to arbitrary involutions.

\section{Symmetric Kullback-Leibler divergence and Jensen-Shannon distance}\label{app:JS}

The bounds on Le Cam's distance allow to compare the Jensen-Shannon (square) distance
\begin{align}
d^2_{JS}(p,q)=\frac{D(p\|\frac{p+q}{2})+D(q\|\frac{p+q}{2})}{2}
\end{align}
with the symmetric Kullback-Leibler divergence, also called the Jeffreys divergence:
\begin{align}
D(p,q)=\frac{D(p\|q)+D(q\|p)}{2}.
\end{align}
Comparing the lower bound \eqref{eq:JSvsLeCam} on Le Cam's distance with the upper bound \eqref{eq:lecamleqtanh}, we conclude that:
\begin{align}
e^{d^2_{JS}(p,q)}-1 \leq \tanh \frac{D(p,q)}{2}.
\end{align}
Using $ \tanh x = 2/(1+e^{-2x})-1$, this can be rewritten as 

\begin{align}
d^2_{JS}(p,q) \leq \ln  \frac{2}{1+e^{-D(p,q)}}
\end{align}
which is the main result of \cite{crooks2008inequalities}. 

\section{Switching systems and bang-bang controls}\label{app:switch}

When controlling the parameters of a dynamics in order to elicit a certain behaviour, it is common to adopt a strategy of abrupt switching (at well-chosen time instants) between two or several continuous-time dynamics. For instance we can consider a time-varying Markov chain which switches between two transition rate matrices.  
One reason for such a switching control strategy is the simplicity of implementation, where one has to simply switch on or off a voltage source, a magnetic flux, a reactant influx, a heater, etc.

A deeper reason is to be found in optimal control theory. Even if we can modulate freely the external force with a real number $u(t)$, some various fundamental or practical reasons often constraint the values of $u$ to an interval $[u_{\min}, u_{\max}]$. Assume we want to optimize an objective for the trajectory of the system, for instance the time needed to reach a desired state (or set of states), over all possible input signals $t \mapsto u(t) \in [u_{\min}, u_{\max}]$. Finding this optimal control signal $u(t)$ results in a variational problem whose solution is, in many cases, a so-called \emph{bang-bang control} (or on-off control) switching abruptly between the extremal values $u_{\min}$ and $u_{\max}$ at well-chosen time instants. Typically, minimum-time control problems are well-known to often result in bang-bang controls (we refer to textbooks such as \cite{kirk2004optimal} for precise mathematical statements, e.g. the Pontryagin maximum principle). A didactical example is the following: suppose that we want to drive a car from point $A$ (at zero speed) to point $B$ (at zero speed) along a line: how to proceed in minimum time? The answer is accelerate as much as is possible for the car, then at some carefully calculated moment, brake as much as possible so at to arrive at $B$ at zero speed.   
A bit switch is another example: if the input voltage is constrained to be in the interval $[u_{\min}, u_{\max}]$, then the abrupt bang-bang switching is, generically speaking, the fastest way to write a bit into the memory. 
Thus the results illustrated on the bit switch have a more general value for systems operated `as fast as possible', hence in an abrupt switching way.

\section{Distinguishing the arrow of time: low error requires high entropy production}\label{app:arrow}

In \cite{rol15},  the authors consider the following decision problem: we are given a trajectory of a stochastic system, which is presented either in the forward time direction or in the backward time direction (with probability $1/2$), and we must make a guess on which direction is being presented to us. This problem offers another application of the main formula \eqref{eq:eps-sigma}, in which $\Omega$ is regarded as a set of trajectories of a stochastic system. We request that this set is  closed under time-reversal. The trajectories can be on fixed time-interval, or of variable length (determined by a random starting and stopping time).
The two distributions $p$ and $q$ of interest are the forward path probability and the reverse path probability, respectively. 
Now $\Delta \sigma$ is, under some circumstances (for instance a constant, or time-symmetric protocol), the entropy production along the path, or a lower bound on it. The probability of error in the maximum likelihood decision is $\epsilon$. Our formula \eqref{eq:eps-sigma} once again shows that a small probability of error requires a large entropy production. While formally identical to Eq. (9) in \cite{rol15}, \eqref{eq:eps-sigma} holds under very general circumstances, e.g. independently of the choice of a certain stopping criterion and for non-Markovian dynamics.


\section{More classical SLs}\label{app:csl}

\subsection{H\"{o}lder's inequality}

Instead of using the bound \eqref{eq:holder0infty}, one could use as well Cauchy-Schwartz inequality to separate the thermodynamic from the kinetic factors appearing in \eqref{eq:LeCamintegral}. H\"{o}lder's inequality can also be used,
\begin{equation}\label{eq:holder}
\left |\int fg  dt \right | \leq \left (\,\int |f|^k dt\,\right)^{1/k}\,\,\left(\,\int |g|^\ell dt\,\right)^{1/\ell},
\end{equation}
for any  $k, \ell \in [1,+\infty]$ such that $1/k + 1/\ell=1$ and functions $f$, $g$ so that the integrals make sense. In fact, this inequality generalizes both Cauchy-Schwartz inequality ($k=\ell=2$) and \eqref{eq:holder0infty} ($k=\infty$, $\ell=1$).
In this way, every choice of $k,\ell$ leads to a variant classical SL of \eqref{eq:LeCamspeedlimit}. The same remark holds for \eqref{eq:CSLTVintegral}.

\subsection{More distances}

Many distances can be defined on the space of probability distributions, all with their respective applications and interpretations. In addition to Le Cam's distance $d_\text{LeCam}$ and Jensen-Shannon's distance $d_\text{JS}$, we can also define Hellinger's distance,
\begin{align}
d_\text{Hell}(p,q) = \frac{1}{2} \| \sqrt{p} - \sqrt{q} |\|_2,
\end{align}
or Fisher's distance $d_\text{Fisher}$, which is the length in Fisher's information metric sense of the shortest path between two distributions. It turns out that these four distances essentially coincide locally.
Indeed, if $q=p + \Delta p $ with $\Delta p \ll 1$, then
\begin{align}
 d^2_\text{LeCam}(p,q) \approx 2 d^2_\text{JS}(p,q) \approx 2 d^2_\text{Hell}(p,q) \approx \frac{1}{2} d^2_\text{Fisher}(p,q) \approx \frac{1}{4} \sum_{\omega} \frac{\Delta p^2}{p}
\end{align}
Thus one can replace Le Cam's distance $d_\text{LeCam}$ with $\sqrt{2} d_\text{Hell}$, or with $\sqrt{2}d_\text{JS}$, or with $\sqrt{1/2}d_\text{Fisher}$  in SLs \eqref{eq:LeCamintegral},\eqref{eq:LeCamspeedlimit},\eqref{eq:LeCamkinetic}.

\section*{References}

\bibliography{references}

\end{document}